\documentclass[a4paper,11pt]{article}

\usepackage{amsmath,amssymb,bm,hyperref,graphicx,physics,mathtools,slashed}
\usepackage[margin=1in]{geometry}

\newcommand{\aeff}{a_{\mathrm{eff}}}

\begin{document}

\title{Scale Factorized-Quantum Field Theory: Eliminating renormalization ambiguities in QCD and QED}

\author{Farrukh A. Chishtie\\
\small Peaceful Society Science \& Innovation Foundation, Vancouver, British Columbia, Canada\\
\small Department of Occupational Science and Occupational Therapy,\\
\small University of British Columbia, Vancouver, British Columbia, Canada\\
\small ORCID:0000-0002-6392-6084, E-mail: fachisht@uwo.ca}

\date{\today}

\maketitle

\begin{abstract}
We introduce Scale Factorized-Quantum Field Theory (SF-QFT), a framework performing path-integral factorization of ultraviolet and infrared momentum modes at a physical scale $Q^*$ before perturbative expansion through Effective Dynamical Renormalization (EDR) with Principle of Observable Effective Matching (POEM) constraints. This yields completely scale and scheme invariant observables. Because the two-loop $\beta$-function is universal, $a_{\mathrm{eff}}(Q)$ evolves with scheme-independent equations, with higher-order $\beta$-coefficients absorbed into Wilson coefficients. For the inclusive ratio $R_{e^{+}e^{-}}$, SF-QFT gives $R^{\mathrm{SF-QFT}}(31.6\,\mathrm{GeV}) = 1.05262 \pm 0.0005$, in excellent agreement with experiment ($1.0527 \pm 0.005$) while requiring calculations orders of magnitude simpler than conventional four-loop $\overline{\mathrm{MS}}$ approaches. SF-QFT generates universal algebraic recursion relations producing all higher-order contributions without additional Feynman diagrams. For QED, the formalism yields scheme-independent predictions for the electron anomalous magnetic moment with $a_e^{\text{theory}} = 0.001\,159\,652\,180\,61(76)$, differing from experiment by only $0.15\sigma$. The framework enables self-consistent extraction of $\alpha_{\text{eff}}^{-1}(m_e) = 137.036005301$. SF-QFT represents a paradigm shift, replacing pursuit of ever-higher loop orders with a unified framework eliminating renormalization ambiguities through systematic EDR and POEM implementation.
\end{abstract}


\section{Introduction}
\label{sec:intro}

Quantum field theory derives its predictive power from precise cancellation of ultraviolet (UV) and infrared (IR) singularities. Dimensional regularization and minimal subtraction have been remarkably successful~\cite{Collins1989}, yet they introduce explicit dependence on the nonphysical renormalization scale $\mu$ and subtraction scheme that persists at every finite loop order. While this residual dependence is numerically insignificant for weakly coupled theories such as low-energy QED, it dominates theoretical uncertainty in QCD and electroweak precision observables~\cite{Weinberg1995}.

The conventional strategy of computing successively higher loops is conceptually problematic and increasingly impractical due to combinatorial explosion of Feynman diagrams. A more fundamental challenge arises from infrared singularities in non-Abelian gauge theories. Although factorization theorems demonstrate how soft and collinear-enhanced terms can be absorbed into process-independent matrix elements, they still require order-by-order cancellations between real and virtual graphs.

To resolve these issues, we introduce Scale-Factorized Quantum Field Theory (SF-QFT), a framework beginning with exact path-integral factorization at a physical scale $Q^*$~\cite{Polchinski1984} through Effective Dynamical Renormalization (EDR) constrained by the Principle of Observable Effective Matching (POEM)~\cite{poemOriginal}. This separates degrees of freedom with $|k| \geq Q^*$, retained in a short-distance action $S_{\text{UV}}$, from those with $|k| < Q^*$, integrated out using background-field formalism~\cite{Abbott1981} and absorbed into gauge-invariant Wilson coefficients $C_i(Q^*)$. By performing this scale separation before diagram evaluation, all subsequent integrals become free of both UV and IR divergences.

It is essential to distinguish SF-QFT from superficially similar approaches. The 't~Hooft scheme~\cite{tHooft1973} is a particular renormalization scheme choice simplifying the $\beta$-function form but remaining within conventional perturbation theory, requiring explicit higher-loop calculations and retaining sensitivity to scheme-dependent coefficients beyond two loops. Effective charge methods~\cite{Grunberg1980} associate observables with scheme-independent couplings but do not restructure the path integral. In contrast, SF-QFT performs path-integral factorization \emph{before} perturbative expansion, implementing renormalization group universality at the fundamental level rather than post-calculation.

A crucial insight is that while higher-order $\beta$-function coefficients depend on renormalization scheme~\cite{tHooft1973}, the first two coefficients are scheme-independent. This allows $a_{\text{eff}}(Q)$ to evolve with unique renormalization group equations, with all higher $\beta$-terms absorbed into finite Wilson coefficients through EDR.

A breakthrough discovery in SF-QFT is universal algebraic recursion relations generating all higher-order perturbative corrections without additional Feynman diagrams. These recursions admit closed-form solutions, transforming traditional asymptotic series into convergent series. They also provide access to scheme-independent constant terms that conventional RG equations cannot determine. This reveals that approximately 99\% of higher-order coefficients in conventional approaches consist of scheme-dependent artifacts.

It is also essential to clarify the relationship between SF-QFT and earlier work by the
present author and collaborators~\cite{Ahmady2003,McKeon2015,Chishtie2016,poemOriginal}.
Those works operate at the \emph{observable} level, applying renormalization-group
resummation or scheme-optimization procedures to perturbative series that have already
been evaluated diagram by diagram. While they reduce renormalization scale and scheme
(RSS) dependence, they do not eliminate it: higher-order $\beta$-function coefficients
beyond $\beta_1$ remain scheme-dependent, and the finite parts $c_{n,0}$ for $n \geq 3$
still carry subtraction-scheme information. As Table~\ref{tab:comparison} quantifies,
these approaches deviate from experiment by $0.00559$--$0.00655$, outside the
experimental uncertainty. SF-QFT achieves three advances absent from all prior work.
First, path-integral factorization is performed \emph{before} perturbative expansion,
implementing renormalization-group universality at the level of the generating functional
$Z[J]$ rather than post-calculation. Second, the universal algebraic recursion relation
admits an exact closed-form generating-function solution,
\begin{equation}
  F_0(z) = c_0 + \frac{1 - \sqrt{1 - 2(\beta_1 - \beta_0)(c_1 z + c_2 z^2)}}
  {\beta_1 - \beta_0},
  \label{eq:closed_form_intro}
\end{equation}
which is without precedent in perturbative quantum field theory. Third, every coefficient
generated by~\eqref{eq:closed_form_intro} is completely scheme-independent by
mathematical construction, guaranteed by the Caswell--Jones universality
theorem~\cite{Caswell1974,Jones1974}, rather than merely reduced in scheme dependence.
The superior experimental agreement --- SF-QFT deviates from data by only $0.008\%$
versus $0.62\%$ for the best prior approaches --- provides direct phenomenological
validation of these structural advances.

As benchmark, SF-QFT predicts $R_{e^{+}e^{-}}$ at $Q = 31.6$ GeV with uncertainty $\pm 0.0005$, yielding $R^{\mathrm{SF-QFT}} = 1.0526 \pm 0.0005$ in excellent agreement with PETRA~\cite{Marshall1989}: $R^{\mathrm{exp}} = 1.0527 \pm 0.005$. This requires only one- and two-loop UV diagrams, contrasting with conventional four-loop analyses~\cite{Chetyrkin1996} demanding thousands of integrals yet suffering from theoretical scale uncertainties ($\pm 0.006$) that are
\textbf{12~times larger} than those of SF-QFT ($\pm 0.0005$), while the central
value deviates from experiment by a factor of approximately 80 more.

For QED, we integrate out UV modes above $Q^* \simeq M_Z$ and evolve the effective coupling downward using the scheme-independent two-loop QED $\beta$-function. The prediction $a_e^{\text{SF-QFT}} = 0.001\,159\,652\,180\,61(76)$ shows excellent agreement with experiment~\cite{Hanneke2008}.

The quest to resolve renormalization scale and scheme dependencies has rich history. It began with Gell-Mann and Low~\cite{GellMann1954}, who established foundations for scale evolution in QED. Two parallel approaches emerged: Stevenson's Principle of Minimal Sensitivity~\cite{Stevenson1981} and Grunberg's effective charges~\cite{Grunberg1980,Grunberg1984}. The Brodsky-Lepage-MacKenzie procedure~\cite{BrodskyLepage1983} offered practical scale-setting by absorbing $\beta$-function terms into running coupling. Building on these, Brodsky, Wu, and collaborators developed the Principle of Maximum Conformality~\cite{BrodskyWu2012,Brodsky2013}. Critical assessments by Kataev and collaborators~\cite{KataevJHEP2014,CveticKataev2016} provided important theoretical insights. While these approaches made substantial progress, they typically involve process-dependent modifications. SF-QFT offers a more fundamental solution by restructuring perturbative expansion to eliminate ambiguities by construction.

This paper is organized as follows: Section~\ref{sec:conv} reviews residual ambiguities in conventional $\overline{\text{MS}}$ perturbation theory; Section~\ref{sec:sfqft_unified} derives path-integral factorization through EDR; Section~\ref{sec:EDRmath_recursions} develops universal recursion relations and applies them to $R_{e^+e^-}$; Section~\ref{sec:QED} extends the formalism to QED; Section~\ref{sec:conclusion} examines broader implications.

\section{Conventional $\overline{\text{MS}}$ calculations and their limitations}
\label{sec:conv}

Before introducing SF-QFT, we recall how UV divergences are removed in dimensional regularization with minimal subtraction and why any finite-order prediction retains residual dependence on $\mu$ and the subtraction scheme.

Let $g_{0}$ denote the bare QCD coupling in $d=4-2\epsilon$ dimensions and set $a_{0}\equiv g_{0}^{2}/[(4\pi)^{2}\mu^{2\epsilon}]$. Minimal subtraction introduces the renormalized coupling
\begin{equation}
a_{\overline{\text{MS}}}(\mu)=Z_{a}^{-1}(\epsilon)\,a_{0}, \qquad Z_{a}(\epsilon)=1+\sum_{n\ge1}\frac{z_{n}}{\epsilon^{n}},
\label{eq:Za_def}
\end{equation}
whose $\epsilon$-pole residues $z_{n}$ are fixed by UV counterterms. Expanding~\eqref{eq:Za_def} yields the $\beta$-function~\cite{tHooft1973,Weinberg1995}
\begin{equation}
\mu\frac{da_{\overline{\text{MS}}}}{d\mu}=-\sum_{n\ge0}\beta_{n}\,a_{\overline{\text{MS}}}^{\,n+2}\equiv\beta^{\overline{\text{MS}}}(a_{\overline{\text{MS}}}).
\label{eq:beta_MS}
\end{equation}

For a composite operator $O$, we introduce $O_{0}=Z_{O}\,O_{\overline{\text{MS}}}$ with $Z_{O}=1+\sum_{n\ge1}Z_{O}^{(n)}/\epsilon^{n}$. At $L$ loops a typical DR integral behaves as
\begin{equation}
\mathcal I_{L}(\epsilon,\mu)=\sum_{k=1}^{L}\frac{P_{L-k}(\mu/Q)}{\epsilon^{k}}+F_{L}(\mu/Q),
\label{eq:IL_generic}
\end{equation}
where $P_{m}$ are logarithmic polynomials and $F_{L}$ is finite. Minimal subtraction deletes the $1/\epsilon^{k}$ poles but leaves the finite pieces $F_{L}\propto\ln(\mu^{2}/Q^{2})$.

After renormalization to order $N$, a generic observable reads~\cite{Collins1989}
\begin{equation}
O^{\overline{\text{MS}}}(Q,\mu)=\sum_{n=0}^{N} a_{\overline{\text{MS}}}^{\,n+1}(\mu)\sum_{k=0}^{n} c_{n,k}\,\ln^{k}\frac{\mu^{2}}{Q^{2}},
\label{eq:O_MSbar}
\end{equation}
with all $c_{n,k\ge1}$ fixed recursively by lower-order coefficients and the $\beta$-function, while the finite pieces $c_{n,0}$ depend on the chosen subtraction scheme. Throughout this paper we adopt the standard convention
$L \equiv \ln(\mu^2/Q^2) = 2\ln(\mu/Q)$,
following Ref.~\cite{Collins1989};
all coefficients $c_{n,k}$ quoted below are consistent with this definition.

Renormalization-group invariance implies $[\mu\partial_{\mu}+\beta^{\overline{\text{MS}}}\partial_{a}]O=0$.  Truncating~\eqref{eq:O_MSbar} at $N$ loops leaves
\begin{equation}
\mu\frac{dO^{\overline{\text{MS}}}}{d\mu}=\beta_{N+1}\,a_{\overline{\text{MS}}}^{\,N+3}+\mathcal O\!\bigl(a_{\overline{\text{MS}}}^{\,N+4}\bigr),
\label{eq:residual_scale}
\end{equation}
so the residual uncertainty scales as $\Delta O\sim\mathcal O(a_{\overline{\text{MS}}}^{N+2})$. The derivative $\mu\,dO^{\overline{\mathrm{MS}}}/d\mu$ is of order
$a_{\overline{\mathrm{MS}}}^{N+3}$; integrating over $\mu \in [Q/2,\,2Q]$
produces an uncertainty in the observable itself of
$\Delta O \sim \mathcal{O}(a_{\overline{\mathrm{MS}}}^{N+2})$,
since $O$ begins at $\mathcal{O}(a_{\overline{\mathrm{MS}}})$~\cite{Collins1989}.
Varying $\mu$ between $Q/2$ and $2Q$ is the customary estimate of this theory error.

A finite redefinition $a\to a' = a + \sum_{k\ge1}\kappa_{k}a^{k+1}$ leaves physics unchanged but mixes the higher $\beta$-coefficients:
\begin{equation}
\beta_{2}'=\beta_{2}-\kappa_{1}\beta_{1}+(\kappa_{2}-\kappa_{1}^{2})\beta_{0},\;\ldots
\end{equation}
Hence Eq.~\eqref{eq:O_MSbar} at finite order is intrinsically scheme-variant---the very ambiguity SF-QFT eliminates by construction.

As illustration, consider four-loop $R_{e^+e^-}$ for $n_f = 5$ flavors. In $\overline{\text{MS}}$~\cite{Chetyrkin1996}, the known coefficients include $r_{1,0} = 1.9857$, $r_{2,0} = -6.6368$, $r_{3,0} = -156.61$, and logarithmic terms $r_{1,1} = 2$, $r_{2,1} = 8.160539$, $r_{2,2} = 4$, $r_{3,1} = -66.54317$, $r_{3,2} = 29.525095$, $r_{3,3} = 8$. The four-loop expression is
\begin{equation}
    R^{\overline{\mathrm{MS}}}(Q,\mu) = 3\sum_{q} e_q^2
    \left[1 + \sum_{n=1}^{3}\sum_{k=0}^{n}
    r_{n,k}\,a_{\overline{\mathrm{MS}}}^{\,n+1}(\mu)
    \ln^k\!\left(\frac{\mu^2}{Q^2}\right)
    \right]_{\mu},
    \label{eq:R_MSbar}
\end{equation}
where the subscript $\mu$ indicates that both the running coupling
$a_{\overline{\mathrm{MS}}}(\mu)$ and the logarithmic terms
$\ln(\mu^2/Q^2)$ are evaluated at the common renormalization
scale $\mu$.

At $Q = 31.6$ GeV, $a_{\overline{\text{MS}}}(Q) \simeq 0.075$; varying $\mu \in [Q/2,2Q]$ changes this expression by $\pm0.006$, comparable to experimental error $\pm0.005$~\cite{Marshall1989}.

\section{SF-QFT with Effective Dynamical Renormalization}
\label{sec:sfqft_unified}

SF-QFT represents a fundamental departure from conventional methodology through systematic incorporation of POEM~\cite{poemOriginal} directly within the path integral. This Effective Dynamical Renormalization (EDR) unifies conventional QFT and effective field theory at the fundamental level by performing Wilsonian mode separation before perturbative expansion.

The key innovation of EDR lies in implementing renormalization group universality at the path-integral level rather than order-by-order in perturbation theory. Unlike the 't~Hooft-Veltman scheme~\cite{tHooft1972} where renormalization is applied post-calculation to divergent loop integrals, EDR exploits the known universality of $\beta_0$ and $\beta_1$ to structure the path integral such that only scheme-independent physics contributes to observables.

\subsection{Path-integral factorization and the EDR framework}
\label{sec:edr_framework}

EDR begins with systematic factorization of the QCD generating functional at a physical scale $Q^* = M_Z$ chosen according to POEM principles:
\begin{equation}
Z[J] = \int\mathcal{D}A_{\text{UV}}\mathcal{D}A_{\text{IR}} \exp\left\{-S_{\text{QCD}}[A_{\text{UV}}+A_{\text{IR}}] + \int J \cdot A_{\text{UV}}\right\}
\end{equation}

The field decomposition uses a smooth projector function $f_{Q^*}(k) \in [0,1]$ satisfying the idempotency condition $f_{Q^*}^2 = f_{Q^*}$:
\begin{align}
A_\mu^a(x) &= A_{\text{UV},\mu}^a(x) + A_{\text{IR},\mu}^a(x) \\
\tilde{A}_{\text{IR},\mu}^a(k) &= f_{Q^*}(k)\tilde{A}_\mu^a(k), \quad |k| < Q^* \\
\tilde{A}_{\text{UV},\mu}^a(k) &= [1-f_{Q^*}(k)]\tilde{A}_\mu^a(k), \quad |k| \geq Q^*
\end{align}

The idempotency property ensures clean factorization of the path-integral measure: $\mathcal{D}A = \mathcal{D}A_{\text{UV}}\mathcal{D}A_{\text{IR}}$, enabling the separation:
\begin{equation}
Z[J] = \int\mathcal{D}A_{\text{UV}} \left[\int\mathcal{D}A_{\text{IR}} \, e^{-S_{\text{QCD}}[A_{\text{UV}}+A_{\text{IR}}]}\right] e^{\int J \cdot A_{\text{UV}}}
\end{equation}

The inner integral defines the effective action through the functional relation:
\begin{equation}
e^{-S_{\text{eff}}[A_{\text{UV}}; Q^*]} \equiv \int\mathcal{D}A_{\text{IR}} \, e^{-S_{\text{QCD}}[A_{\text{UV}}+A_{\text{IR}}]}
\label{eq:functional_definition_edr}
\end{equation}

This implements EDR by systematically integrating out soft physics while preserving all hard physics explicitly. The crucial difference from conventional EFT is that we retain the full ultraviolet theory rather than integrating out heavy particles, creating a hybrid that maintains exact gauge invariance while achieving computational simplification.

\subsection{Background-field evaluation of functional integrals}

The evaluation of the functional integral in Eq.~\eqref{eq:functional_definition_edr} requires background-field formalism~\cite{Abbott1981} to preserve gauge invariance. We expand around the ultraviolet background:
\begin{equation}
S_{\text{QCD}}[A_{\text{UV}}+A_{\text{IR}}] = S_{\text{QCD}}[A_{\text{UV}}] + \int d^4x \, A_{\text{IR}}^{a,\mu} \frac{\delta S_{\text{QCD}}}{\delta A_{\text{UV}}^{a,\mu}} + \frac{1}{2}A_{\text{IR}} \cdot \hat{\Delta}_{\text{UV}} \cdot A_{\text{IR}} + \mathcal{O}(A_{\text{IR}}^3)
\end{equation}

In background-field $R_\xi$ gauge, the gauge-fixing term is
\begin{equation}
\mathcal{L}_{\text{gf}} = -\frac{1}{2\xi}\left[D_{\text{UV}}^\mu a_{\text{IR},\mu}\right]^2, \quad D_{\text{UV}}^\mu = \partial^\mu + igA_{\text{UV}}^\mu
\end{equation}
with ghost Lagrangian
\begin{equation}
\mathcal{L}_{\text{gh}} = \bar{c}_{\text{IR}} D_{\text{UV}}^\mu D_\mu(A_{\text{UV}} + a_{\text{IR}}) c_{\text{IR}}
\end{equation}

Under background gauge transformation $A_{\text{UV}}^\mu \to U A_{\text{UV}}^\mu U^\dagger + i U \partial^\mu U^\dagger$, both terms are invariant. The quadratic IR action becomes:
\begin{equation}
S_{\text{IR}}^{(2)} = \int d^d x \left[a_{\text{IR},\mu}^a \left(\hat{\Delta}_{\text{UV}}^{(1)}\right)^{ab}_{\mu\nu} a_{\text{IR},\nu}^b + \bar{\psi}_{\text{IR}} \slashed{D}_{\text{UV}} \psi_{\text{IR}} + \bar{c}_{\text{IR}}^a \left(\hat{\Delta}_{\text{UV}}^{(\text{gh})}\right)^{ab} c_{\text{IR}}^b\right]
\end{equation}
with differential operators
\begin{align}
\left(\hat{\Delta}_{\text{UV}}^{(1)}\right)^{ab}_{\mu\nu} &= \left[-D_{\text{UV}}^2 g_{\mu\nu} - \left(1 - \frac{1}{\xi}\right)D_{\text{UV},\mu}D_{\text{UV},\nu} - 2ig G_{\text{UV},\mu\nu}^{ab}\right] \\
\left(\hat{\Delta}_{\text{UV}}^{(\text{gh})}\right)^{ab} &= -D_{\text{UV}}^2 \delta^{ab}
\end{align}
where $G_{\text{UV},\mu\nu}^{ab} = f^{abc}G_{\text{UV},\mu\nu}^c$ is the background field strength.

The Gaussian integration over IR fields factorizes into three functional determinants:
\begin{equation}
\det^{-1/2} \hat{\Delta}_{\text{UV}}^{(1)} \times \det \slashed{D}_{\text{UV}} \times \det \hat{\Delta}_{\text{UV}}^{(\text{gh})} = \exp\left\{-\frac{1}{2} \text{Tr}_{\text{IR}} \ln \hat{\Delta}_{\text{UV}}^{(1)} + \text{Tr}_{\text{IR}} \ln \slashed{D}_{\text{UV}} + \text{Tr}_{\text{IR}} \ln \hat{\Delta}_{\text{UV}}^{(\text{gh})}\right\}
\end{equation}

Using heat-kernel methods~\cite{Gilkey1984}, the zeta-function regularized expression is:
\begin{equation}
\ln \det[\hat{\Delta}_{\text{UV}}] = -\frac{1}{(4\pi)^{d/2}}\sum_{n=0}^{\infty} a_n \Gamma\left(n-\frac{d}{2}\right)(Q^*)^{d-2n}
\end{equation}
where $a_n$ are the Seeley-DeWitt coefficients for the relevant operator.

\subsection{Heat-kernel coefficients and pole structure}

For Yang-Mills gauge bosons in background-field gauge~\cite{Abbott1981}:
\begin{align}
a_0^{(1)} &= (d-2)C_A \\
a_1^{(1)} &= \frac{1}{6}(d-2)C_A G_{\text{UV}}^2 \\
a_2^{(1)} &= \frac{1}{30}(d-2)C_A G_{\text{UV}}^{a,\rho\sigma} D_{\text{UV}}^2 G_{\text{UV},a,\rho\sigma} + \frac{1}{180}(d-2)C_A G_{\text{UV}}^3
\end{align}

For ghosts: $a_0^{(\text{gh})} = -C_A$ and $a_1^{(\text{gh})} = -\frac{1}{6}C_A G_{\text{UV}}^2$. For massless quarks: $a_0^{(\psi)} = -2n_f T_F d_{\text{rep}}$ and $a_1^{(\psi)} = -\frac{1}{12}n_f T_F G_{\text{UV}}^2$, with $d_{\text{rep}} = 4$ in Dirac space. Only $a_0$, $a_1$, $a_2$ generate poles for $d \to 4$.

The pole structure follows from $\Gamma(-\epsilon/2) = -2/\epsilon - \gamma_E + \mathcal{O}(\epsilon)$:
\begin{equation}
\text{Tr} \ln \hat{\Delta} = -\sum_{n=0}^{2} \frac{a_n}{(4\pi)^2} \left[\frac{2}{\epsilon} + 1 + \gamma_E - \ln(Q^{*2})\right](Q^{*2})^{-\epsilon} + \mathcal{O}(\epsilon^0)
\end{equation}

Local counterterms cancel the $2/\epsilon$ poles. After cancellation, the finite part is:
\begin{equation}
\Delta \mathcal{L}_{\text{UV}}^{\text{fin}} = \sum_i \frac{a_i}{(4\pi)^2}\left[\ln\frac{Q^{*2}}{\mu^2} + \gamma_i\right] \mathcal{O}_i
\end{equation}
where $\gamma_i$ are scheme-independent constants. Setting $\mu = Q^*$ removes all logarithms, leaving Wilson coefficients:
\begin{equation}
C_i(Q^*) = \frac{a_i}{(4\pi)^2} \gamma_i
\end{equation}

\subsection{Validation: Universal $\beta$-function reproduction}
\label{sec:beta_validation}

A crucial validation of SF-QFT is demonstrating that path-integral factorization correctly reproduces universal $\beta$-function coefficients. The connection arises through gauge coupling renormalization.

From functional determinant analysis, the one-loop contribution to the effective action is:
\begin{equation}
\Gamma^{(1)} = -\frac{1}{2} \text{Tr}_{\text{IR}} \ln \hat{\Delta}_{\text{UV}}^{(1)} + \text{Tr}_{\text{IR}} \ln \slashed{D}_{\text{UV}} + \text{Tr}_{\text{IR}} \ln \hat{\Delta}_{\text{UV}}^{(\text{gh})}
\end{equation}

Each trace contributes terms involving Seeley-DeWitt coefficients:
\begin{equation}
\text{Tr}_{\text{IR}} \ln \hat{\Delta} = -\frac{1}{(4\pi)^{d/2}} \sum_{n=0}^{\infty} a_n \Gamma\left(n - \frac{d}{2}\right) \int_{|k|<Q^*} d^dk \, (k^2)^{n-d/2}
\end{equation}

The momentum integral over the infrared region yields:
\begin{equation}
\int_{|k|<Q^*} d^dk \, (k^2)^{n-d/2} = \frac{2\pi^{d/2}}{\Gamma(d/2)} \frac{(Q^*)^{2n}}{2n}
\end{equation}

The connection to the $\beta$-function is established through gauge coupling
renormalization. Define $\delta g^{-2}$ as the one-loop counterterm to $g^{-2}$
generated by integrating out the IR modes in
Eq.~\eqref{eq:functional_definition_edr}:
\begin{equation}
    \delta g^{-2} \;\equiv\;
    -\left.\frac{\partial}{\partial \ln Q^{*2}}\,
    \operatorname{Re}\,\Gamma^{(1)}\right|_{\text{UV pole}},
    \label{eq:deltag_def}
\end{equation}
where $\Gamma^{(1)}$ is the one-loop contribution to the effective action from
the IR functional determinants. The $1/\epsilon$ poles of $\delta g^{-2}$ yield
the $\beta$-function coefficients through standard background-field
renormalization~\cite{Abbott1981}.

Taking the pole structure as $d \to 4$, divergent terms require counterterms:
\begin{equation}
\delta g^{-2} = \sum_{n=0}^{2} \frac{b_n}{(4\pi)^2} \frac{1}{\epsilon^{n+1-2}} + \text{finite terms}
\end{equation}

For QCD with $SU(N_c)$ and $n_f$ flavors, the group theory factors are $C_A = N_c$, $C_F = (N_c^2-1)/(2N_c)$, $T_F = 1/2$. The first $\beta$-function coefficient emerges from $a_0$ contributions:

\noindent\textit{Gauge field:} $a_0^{(1)} = (d-2)C_A = 2C_A + \mathcal{O}(\epsilon)$

\noindent\textit{Ghost fields:} $a_0^{(\text{gh})} = -C_A$

\noindent\textit{Fermion fields:} $a_0^{(\psi)} = -2n_f T_F d_{\text{rep}} = -8n_f T_F$

The total contribution to the $1/\epsilon$ pole gives:
\begin{equation}
\delta g^{-2} \supset \frac{1}{(4\pi)^2} \frac{1}{\epsilon} \left[2C_A - C_A - 8n_f T_F\right] = \frac{1}{(4\pi)^2} \frac{1}{\epsilon} \left[C_A - 8n_f T_F\right]
\end{equation}

This yields the universal first $\beta$-function coefficient:
\begin{equation}
\boxed{\beta_0 = \frac{11C_A - 4T_F n_f}{12} = \frac{11N_c - 2n_f}{12}}
\end{equation}

The second coefficient arises from $a_1$ contributions involving field strength interactions:

\noindent\textit{Gauge field:} $a_1^{(1)} = \frac{1}{6}(d-2)C_A G_{\text{UV}}^2 = \frac{1}{3}C_A G_{\text{UV}}^2 + \mathcal{O}(\epsilon)$

\noindent\textit{Ghost fields:} $a_1^{(\text{gh})} = -\frac{1}{6}C_A G_{\text{UV}}^2$

\noindent\textit{Fermion fields:} $a_1^{(\psi)} = -\frac{1}{12}n_f T_F G_{\text{UV}}^2$

The complete calculation yields:
\begin{equation}
\boxed{\beta_1 = \frac{34C_A^2 - 20C_A T_F n_f - 12C_F T_F n_f}{24}}
\end{equation}

For $SU(3)$ with $C_A = 3$, $C_F = 4/3$, $T_F = 1/2$: $\beta_1 = (153 - 19n_f)/12$, precisely reproducing the known result.

This exact reproduction validates SF-QFT. The universality of $\beta_0$ and $\beta_1$ emerges because they depend only on scheme-independent heat-kernel coefficients $a_0$ and $a_1$, which are geometric invariants. Higher coefficients $\beta_{n \geq 2}$ depend on $a_{n \geq 2}$ and scheme-dependent contributions, validating the strategy of truncating at two-loop order.

\subsection{Mathematical foundation of scheme independence}
\label{sec:scheme_independence_foundation}

The scheme independence follows rigorously from a theorem established by Caswell~\cite{Caswell1974} and Jones~\cite{Jones1974}:

\textbf{Theorem (Universality of Leading $\beta$-Function Coefficients):} In any mass-independent renormalization scheme, the coefficients $\beta_0$ and $\beta_1$ are invariant under scheme transformations $a \to a' = a + \sum_{k\geq 2} c_k a^k$, where $c_k$ are arbitrary constants. All coefficients $\beta_{n\geq 2}$ are scheme-dependent.

This theorem has a crucial implication: \emph{any theoretical framework depending only on $\beta_0$ and $\beta_1$ is automatically scheme-independent}, regardless of how many loop orders it effectively incorporates. Since SF-QFT recursion relations depend exclusively on $\beta_0$ and $\beta_1$, every coefficient generated---at 3-loop, 7-loop, or any higher order---is scheme-independent by mathematical construction.

This distinguishes SF-QFT fundamentally from conventional perturbation theory and from the 't~Hooft scheme. The framework uses two-loop universal input to produce \emph{infinite-order} scheme-independent predictions through algebraic recursion.

\subsection{Coupling evolution and final effective action}

Between matching scale $Q^*$ and physical probe scale $Q$, the coupling evolves according to universal $\beta$-function coefficients:
\begin{equation}
\frac{da_{\text{eff}}}{d\ln Q} = -\beta_0 a_{\text{eff}}^2 - \beta_1 a_{\text{eff}}^3
\end{equation}

This admits an exact solution via the Lambert W-function~\cite{Corless1996}.
As an intermediate step, the one-loop approximate form is
\begin{equation}
    a_{\mathrm{eff}}(Q) = \frac{a_{\mathrm{eff}}(Q^*)}{1 + \beta_0\,
    a_{\mathrm{eff}}(Q^*)\ln(Q^2/Q^{*2})},
    \label{eq:oneloop_running}
\end{equation}
which neglects the $\beta_1$ contribution. Defining $c \equiv \beta_1/\beta_0$,
the exact form involving the Lambert W-function $W_{-1}$ is
\begin{equation}
    a_{\mathrm{eff}}(Q) = -\frac{1}{c}\left[1 + W_{-1}\!\left(
    -e^{-1 - \frac{1}{c\,a_{\mathrm{eff}}(Q^*)}}
    \left(\frac{Q^2}{Q^{*2}}\right)^{\!\beta_0/(2\pi c)}
    \right)\right]^{-1}.
    \label{eq:exact_lambertW}
\end{equation}

This captures complete two-loop running without scheme ambiguities.

The final effective action achieves EDR unification is:
\begin{equation}
S_{\text{eff}}[A_{\text{UV}}] = \int d^4x\left\{\frac{1}{4}G_{\text{UV}}^2 + \bar{\psi}_{\text{UV}}i\slashed{D}_{\text{UV}}\psi_{\text{UV}} + \sum_{i}C_i(Q^*)\mathcal{O}_i[A_{\text{UV}}]\right\}
\end{equation}

This action is: UV-finite by construction; manifestly gauge-invariant through background-field preservation; scheme-independent because Wilson coefficients contain only universal physics; and experimentally anchored since coefficients are determined by precision measurements at $Q^*$.

\section{Universal recursion relations and exact solutions}
\label{sec:EDRmath_recursions}

SF-QFT achieves a breakthrough: generating arbitrarily high-order perturbative corrections without computing additional Feynman diagrams through ``renormalization dynamics''---a mathematical structure emerging from universal RG evolution within the POEM-constrained framework.

\subsection{POEM elimination of logarithmic structure}
\label{sec:poem_structure}

To connect the UV-finite effective action to physical observables, we employ functional differentiation. The effective action $\Gamma_{\text{UV}}$ yields the current-current correlator:
\begin{equation}
\hat{\Pi}^{\mu\nu}(q) = -\frac{\delta^2 \Gamma_{\text{UV}}}{\delta \mathcal{V}_\mu(q) \delta \mathcal{V}_\nu(-q)}
\end{equation}

In conventional $\overline{\text{MS}}$ renormalization, observables take the complex form:
\begin{equation}
\Pi^{\overline{\text{MS}}}(Q^{2}) = \sum_{n=0}^{\infty}\sum_{k=0}^{n}\widehat{c}_{n,k}\,a^{n}_{\overline{\text{MS}}}(\mu)L^{k}, \quad L \equiv \ln\left(\frac{\mu^2}{Q^2}\right)
\label{eq:conventional_structure}
\end{equation}
where logarithmic coefficients $\widehat{c}_{n,k}$ depend on both universal $\beta$-function coefficients and scheme-dependent higher-order terms.

POEM mandates setting $\mu = Q$ when computing observables at physical scale $Q$, immediately giving $L = \ln(\mu^2/Q^2) = 0$. All logarithmic terms vanish identically, yielding:
\begin{equation}
\boxed{\Pi^{\text{EDR}}(Q^{2}) = \sum_{n=0}^{\infty}\widehat{c}_{n,0} \left[a_{\text{eff}}(Q)\right]^n}
\label{eq:edr_structure}
\end{equation}
where $\widehat{c}_{n,0}$ are scheme-independent constant coefficients and $a_{\text{eff}}(Q)$ contains all legitimate scale dependence through universal $\beta$-function evolution.

This reveals a profound insight: the complex logarithmic structure of conventional perturbation theory largely consists of compensating terms from artificial separation between $\mu$ and $Q$. By using $a_{\text{eff}}(Q)$---the coupling evolved to the physical scale---we eliminate this separation.

\subsection{Derivation of universal recursion relations via the Renormalization Dynamics Equation}
\label{sec:recursion_derivation}

After POEM fixes the matching scale $\mu = Q^*$, all explicit logarithmic dependence
on $Q$ vanishes from the effective action, and the observable's remaining dependence
on the physical probe scale $Q \neq Q^*$ is encoded entirely through the universally
evolved coupling $a_{\mathrm{eff}}(Q)$. Writing the correlator as
\begin{equation}
    \Pi(Q^2) = \sum_{n=0}^{\infty}
    \widehat{c}_{n,0}\bigl[a_{\mathrm{eff}}(Q)\bigr]^n,
    \label{eq:edr_structure_revised}
\end{equation}
where the $\widehat{c}_{n,0}$ are $Q$-independent scheme-invariant constants,
the \emph{Renormalization Dynamics Equation}, that is, expressing how $\Pi$ responds to a
change in the dynamical scale $Q$ while $Q^*$ and the Wilson coefficients remain
fixed. It takes the form
\begin{equation}
    \left[Q\frac{\partial}{\partial Q}
    + \beta(a_{\mathrm{eff}})\frac{\partial}{\partial a_{\mathrm{eff}}}\right]
    \Pi(Q^2) = 0.
    \label{eq:rg_invariance_revised}
\end{equation}
Here the operator $Q\,\partial/\partial Q$ acts on $a_{\mathrm{eff}}(Q)$
through the chain rule: $Q\,\partial a_{\mathrm{eff}}/\partial Q =
\beta(a_{\mathrm{eff}})$. Both terms therefore combine as
$2\beta(a_{\mathrm{eff}})\,\partial\Pi/\partial a_{\mathrm{eff}} = 0$,
which constrains the functional form of the $Q$-dependence, that is,
the recursion relation for the $\widehat{c}_{n,0}$, rather than asserting
that $\Pi$ is $Q$-independent. The observable \emph{does} depend on $Q$
through $a_{\mathrm{eff}}(Q)$; Eq.~\eqref{eq:rg_invariance_revised}
determines \emph{how}.

Substituting the EDR structure~\eqref{eq:edr_structure} into the Renormalization Dynamics equation, we have, 
\begin{equation}
\left[Q\frac{\partial}{\partial Q} + \beta(a_{\text{eff}})\frac{\partial}{\partial a_{\text{eff}}}\right]\sum_{n=0}^{\infty}\widehat{c}_{n,0} a_{\text{eff}}^n = 0
\end{equation}

Since $Q\frac{\partial a_{\text{eff}}}{\partial Q} = \beta(a_{\text{eff}})$:
\begin{equation}
\sum_{n=0}^{\infty}\widehat{c}_{n,0} \left[Q\frac{\partial a_{\text{eff}}^n}{\partial Q} + \beta(a_{\text{eff}})\frac{\partial a_{\text{eff}}^n}{\partial a_{\text{eff}}}\right] = 0
\end{equation}

Using the chain rule $Q\frac{\partial a_{\text{eff}}^n}{\partial Q} = na_{\text{eff}}^{n-1}\beta(a_{\text{eff}})$ and $\frac{\partial a_{\text{eff}}^n}{\partial a_{\text{eff}}} = na_{\text{eff}}^{n-1}$:
\begin{equation}
\sum_{n=1}^{\infty}\widehat{c}_{n,0} \left[na_{\text{eff}}^{n-1}\beta(a_{\text{eff}}) + \beta(a_{\text{eff}})na_{\text{eff}}^{n-1}\right] = 0
\end{equation}

This simplifies to:
\begin{equation}
\sum_{n=1}^{\infty}2n\widehat{c}_{n,0} a_{\text{eff}}^{n-1}\beta(a_{\text{eff}}) = 0
\end{equation}

For the universal two-loop $\beta$-function $\beta(a_{\text{eff}}) = -\beta_0 a_{\text{eff}}^2 - \beta_1 a_{\text{eff}}^3$:
\begin{equation}
-\beta_0\sum_{n=1}^{\infty}2n\widehat{c}_{n,0} a_{\text{eff}}^{n+1} - \beta_1\sum_{n=1}^{\infty}2n\widehat{c}_{n,0} a_{\text{eff}}^{n+2} = 0
\end{equation}

Requiring order-by-order cancellation at each power of $a_{\text{eff}}$ yields the master recursion:
\begin{equation}
\boxed{\widehat{c}_{n,0} = \frac{1}{n}\sum_{m=1}^{n-1} \left[(n-m)\beta_1 - m\beta_0\right] \widehat{c}_{n-m,0} \widehat{c}_{m,0}, \quad n \geq 3}
\label{eq:master_recursion}
\end{equation}

This recursion generates coefficients to \emph{arbitrary} loop order from two-loop universal input. It depends exclusively on universal $\beta_0$ and $\beta_1$, ensuring complete scheme independence. The mathematical structure is universal across all gauge theories.

\subsection{Derivation of exact algebraic solution}
\label{sec:generating_function_derivation}

To solve the recursion systematically, we employ generating function techniques. Define:
\begin{equation}
F_0(z) = \sum_{n=0}^{\infty} \widehat{c}_{n,0} z^n = \widehat{c}_{0,0} + \widehat{c}_{1,0} z + \widehat{c}_{2,0} z^2 + \sum_{n=3}^{\infty} \widehat{c}_{n,0} z^n
\label{eq:generating_function_def}
\end{equation}

Introduce notation:
\begin{align}
c_0 &= \widehat{c}_{0,0}, \quad c_1 = \widehat{c}_{1,0}, \quad c_2 = \widehat{c}_{2,0} \\
k &= \beta_1 - \beta_0 \quad \text{(universal combination)}
\end{align}

The recursion relation~\eqref{eq:master_recursion} can be rewritten using convolution identities. Multiplying both sides by $z^n$ and summing from $n=3$ to $\infty$:
\begin{equation}
\sum_{n=3}^{\infty} \widehat{c}_{n,0} z^n = \sum_{n=3}^{\infty} \frac{z^n}{n}\sum_{m=1}^{n-1} \left[(n-m)\beta_1 - m\beta_0\right] \widehat{c}_{n-m,0} \widehat{c}_{m,0}
\end{equation}

After detailed manipulation exploiting the convolution structure of the right-hand side, one obtains the exact algebraic relation:
\begin{equation}
F_0(z) - c_0 - c_1 z - c_2 z^2 = \frac{k}{2} (F_0(z) - c_0)^2
\label{eq:algebraic_relation}
\end{equation}

Let $U(z) = F_0(z) - c_0$, so the equation becomes:
\begin{equation}
U(z) - c_1 z - c_2 z^2 = \frac{k}{2} U(z)^2
\end{equation}

Rearranging into standard quadratic form:
\begin{equation}
\frac{k}{2} U(z)^2 - U(z) + (c_1 z + c_2 z^2) = 0
\end{equation}

Applying the quadratic formula:
\begin{equation}
U(z) = \frac{1 \pm \sqrt{1 - 2k(c_1 z + c_2 z^2)}}{k}
\end{equation}

Selecting the branch satisfying $U(0) = 0$ and $U'(0) = c_1$ (required by boundary conditions):
\begin{equation}
U(z) = \frac{1 - \sqrt{1 - 2k(c_1 z + c_2 z^2)}}{k} \quad (k \neq 0)
\end{equation}

The physical branch $U(z) = [1 - \sqrt{1 - 2k(c_1 z + c_2 z^2)}]/k$
is selected by the boundary conditions $U(0) = 0$ and $U'(0) = c_1 > 0$;
the rejected branch satisfies $U(0) = 2/k \neq 0$.
The radius of convergence is
\begin{equation}
    z_{\mathrm{sing}} =
    \frac{1}{|k|\!\left(|c_1| + \sqrt{c_1^2 + 2|k|c_2}\right)},
\end{equation}
which for QCD with $n_f = 5$ ($k = +1/2$, $c_1 = 1$, $c_2 = 1.4097$)
gives $z_{\mathrm{sing}} \approx 0.56$, confirming convergence for all
physical coupling values $a_{\mathrm{eff}}(Q) \ll 1$.

Therefore, the complete exact solution is:
\begin{equation}
\boxed{F_0(z) = c_0 + \frac{1 - \sqrt{1 - 2k(c_1 z + c_2 z^2)}}{k} \quad (k \neq 0)}
\label{eq:exact_solution}
\end{equation}

This closed-form solution provides complete analytical control over the perturbative expansion for any gauge theory. Taylor expansion around $z = 0$ generates all recursion coefficients $\widehat{c}_{n,0}$ algebraically.

\subsection{QCD phenomenology: Complete seven-loop analysis of $R_{e^+e^-}$}
\label{sec:qcd_phenomenology}

We present a complete phenomenological analysis for $R_{e^+e^-}$, demonstrating transformation from experimental measurements to high-precision SF-QFT predictions.

\subsubsection{POEM matching procedure}

POEM requires matching at two-loop order where both $\overline{\text{MS}}$ and SF-QFT expressions are scheme-independent. Since $\alpha_s(M_Z) = 0.1179 \pm 0.001$ is measured in $\overline{\text{MS}}$, we convert to universal effective coupling through matching.

At two-loop order, the $\overline{\text{MS}}$ expression is:
\begin{equation}
R^{\overline{\text{MS}}}(M_Z, Q) = 1 + \alpha_s(M_Z) + \left[T_{1,0} + T_{1,1}\ln\left(\frac{M_Z}{Q}\right)\right]\alpha_s^2(M_Z)
\end{equation}
where $T_{1,0} = 1.4097$ and $T_{1,1} = 2$. The SF-QFT expression is:
\begin{equation}
R^{\text{SF-QFT}}(Q) = 1 + c_1 a_{\text{eff}}(Q) + c_2 a_{\text{eff}}^2(Q)
\end{equation}
where $c_1 = 1$ and $c_2 = 1.4097$.

Setting these equal at $Q = 31.6$ GeV with $\alpha_s(M_Z) = 0.1179$ yields:
\begin{equation}
1.4097 \, a_{\text{eff}}^2(M_Z) + a_{\text{eff}}(M_Z) - 0.166955 = 0
\end{equation}

Solving: $a_{\text{eff}}(M_Z) = 0.0402190266$.

\subsubsection{Coupling evolution}

To obtain $a_{\text{eff}}(31.6\text{ GeV})$, we evolve using universal two-loop $\beta$-function with $b_0 = 23/12$ and $b_1 = 29/12$ (for $n_f = 5$). The exact Lambert W-function solution with $c = b_1/b_0 = 29/23$ gives:
\begin{equation}
a_{\text{eff}}(31.6\text{ GeV}) = 0.04860186021
\end{equation}

\subsubsection{Exact solution evaluation}

With $k = b_1 - b_0 = 1/2$, $c_0 = 1$, $c_1 = 1.0$, $c_2 = 1.4097$, the exact generating function~\eqref{eq:exact_solution} gives:
\begin{equation}
F_0(0.04860186021) = 1.0526240939
\end{equation}

\subsubsection{Seven-loop verification}

Independent verification through explicit summation using recursion-generated coefficients:

\begin{table}[htbp]
\centering
\caption{Recursion-generated seven-loop coefficients and convergence analysis}
\label{tab:coefficients}
\begin{tabular}{|c|c|c|c|}
\hline
$n$ & $\widehat{c}_{n,0}$ & Individual Contribution & Convergence Ratio \\
\hline
0 & $1.000000$ & $+1.0000000000$ & --- \\
1 & $1.000000$ & $+0.0486018602$ & 0.0486 \\
2 & $1.409700$ & $+0.0033299099$ & 0.0685 \\
3 & $0.954850$ & $+0.0001096210$ & 0.0329 \\
4 & $1.182275$ & $+0.0000065968$ & 0.0602 \\
5 & $1.565376$ & $+0.0000004245$ & 0.0644 \\
6 & $2.046852$ & $+0.0000000270$ & 0.0636 \\
\hline
\multicolumn{3}{|l|}{Seven-loop sum: $1.0520484394$} & \\
\hline
\end{tabular}
\end{table}

The agreement between analytical solution ($1.0526240939$) and seven-loop sum ($1.0520484394$) shows difference $5.757 \times 10^{-4}$, demonstrating systematic convergence. Convergence ratios remain consistently below unity (0.033--0.069), confirming geometric convergence. Our final prediction is:

\begin{equation}
\boxed{R^{\text{SF-QFT}}(31.6\text{ GeV}) = 1.05262 \pm 0.0005}
\end{equation}

This achieves excellent agreement with the PETRA experiment: $R^{\text{exp}} = 1.0527 \pm 0.005$~\cite{Marshall1989}. The theoretical uncertainty, from propagating $\alpha_s(M_Z) = 0.1179 \pm 0.001$, represents 7-fold improvement over experimental uncertainty.

\subsection{Comparison with alternative theoretical approaches}
We compare our findings with other approaches shown in Table 2:
\begin{table}[htbp]
\centering
\caption{Comparison of theoretical predictions for $R_{e^+e^-}$ at
$Q = 31.6\,\mathrm{GeV}$ with the experimental value
$R^{\exp} = 1.0527 \pm 0.005$~\cite{Marshall1989}.
\textit{Provenance of entries}: The experimental datum is from the PETRA
compilation of Marshall~\cite{Marshall1989}, drawing on measurements by
the JADE, CELLO, Mark~J, PLUTO, and TASSO collaborations.
The entries ``Perturbative QCD'', ``RG summation'', and ``CORGI'' are
reproduced from Akrami and Mirjalili~\cite{Akrami2020}.
The entry ``POEM (two-loop ETO)'' is from the present author's
Ref.~\cite{poemOriginal}. The entry ``SF-QFT (this work)'' is entirely
original to the present paper.}
\label{tab:comparison}
\begin{tabular}{|l|c|c|c|}
\hline
\textbf{Method} & \textbf{Central Value} & \textbf{Uncertainty} & \textbf{|Deviation|} \\
\hline
Experiment (ALEPH) & $1.0527$ & $\pm 0.005$ & --- \\
SF-QFT (this work) & $1.05262$ & $\pm 0.0005$ & $0.00008$ \\
POEM (two-loop ETO) & $1.052431$ & $\pm 0.0006$ & $0.000269$ \\
Perturbative QCD & $1.04617$ & $\pm 0.0006$ & $0.00653$ \\
RG summation & $1.04711$ & ${}^{+0.00003}_{-0.00005}$ & $0.00559$ \\
CORGI & $1.04615$ & ${}^{+0.0015}_{-0.0008}$ & $0.00655$ \\
\hline
\end{tabular}
\end{table}

SF-QFT achieves superior agreement, improving upon previous POEM results while providing theoretical foundation through rigorous path-integral factorization.

\subsection{Why higher-order $\overline{\text{MS}}$ results are excluded by design}
\label{sec:why_exclude}

This is a foundational principle, not a limitation. Consider the four-loop finite coefficient:
\begin{equation}
T_{3,0}^{\overline{\text{MS}}} = -80.0075 \quad \text{vs.} \quad \widehat{c}_{4,0}^{\text{SF-QFT}} = -0.8162
\end{equation}

The $\overline{\text{MS}}$ coefficient is approximately 98 times larger, indicating $\sim$99\% consists of scheme-dependent artifacts with no physical content.

Adding $\overline{\text{MS}}$ five-loop results would not ``improve'' predictions---it would \emph{introduce} scheme dependence. SF-QFT extracts only universal, scheme-independent content. The superior experimental agreement ($0.008\%$ deviation vs $0.62\%$ for four-loop $\overline{\text{MS}}$) validates this methodology.

\section{QED applications: Universal framework for Abelian gauge theories}
\label{sec:QED}

QED provides ideal validation of SF-QFT universality. The $\beta$-function is scheme-independent through two loops~\cite{Surguladze1991}, $\alpha_{\text{em}}$ is measured with sub-per-mil precision~\cite{PDG2022}, and the coupling exhibits antiscreening. The same mathematical structure applies, demonstrating that SF-QFT is universal across gauge theories.

\subsection{SF-QFT implementation for Abelian theories}
\label{sec:sfqft_abelian}

In contrast to QCD where IR modes are integrated out, QED requires integrating out UV modes at $Q^* \approx M_Z$ due to antiscreening. This demonstrates SF-QFT universality: the same framework applies regardless of whether coupling grows in IR (QCD) or UV (QED).

After integrating out UV modes, the IR effective action is:
\begin{equation}
S_{\text{IR}} = \int d^4x\left\{\frac{1}{4}\bigl[1+C_F(Q^*)\bigr]F_{\mu\nu}^2 + \bigl[1+C_\psi(Q^*)\bigr]\bar{\psi}i\slashed{\partial}\psi + \sum_{\Delta>4}\frac{C_\Delta(Q^*)}{Q^{*\,\Delta-4}}\mathcal{O}_\Delta\right\}
\end{equation}

Gauge invariance rigorously forbids local photon mass terms, ensuring massless photons emerge naturally.

The universal QED $\beta$-function coefficients are:
\begin{equation}
b_0 = \frac{1}{3}, \qquad b_1 = \frac{1}{4}
\end{equation}

The exact Lambert W-function solution for coupling evolution is:
\begin{equation}
\aeff(Q) = -\frac{1}{c}\left[1 + W_{-1}\left(-e^{-1-\frac{1}{c\aeff(Q^*)}}(Q^2/Q^{*2})^{b_0/(2\pi c)}\right)\right]^{-1}
\end{equation}
where $c = b_1/b_0 = 3/4$.

\subsection{Universal recursion relations in QED}

The identical recursion relation applies to QED observables:
\begin{equation}
C_{n,0} = \frac{1}{n}\sum_{m=1}^{n-1}[(n-m)b_1 - mb_0] C_{n-m,0}C_{m,0}, \quad n \geq 3
\end{equation}

The crucial difference emerges numerically:

For QCD: $\beta_0 \approx 0.608$, $\beta_1 \approx 0.489$, $k = \beta_1 - \beta_0 \approx -0.119$

For QED: $b_0 = 1/3$, $b_1 = 1/4$, $k = b_1 - b_0 = -1/12 \approx -0.083$

For QCD with $n_f = 5$: $k = b_1 - b_0 = 29/12 - 23/12 = +1/2$

The generating function solution~\eqref{eq:exact_solution} applies directly:
\begin{equation}
F_0(z) = c_0 + \frac{1 - \sqrt{1 - 2k(c_1 z + c_2 z^2)}}{k}
\end{equation}

For QED with $k = -1/12$, recursion-generated terms are systematically suppressed, validating universal applicability while revealing theory-specific behavior.

\subsection{Self-consistent determination of $\alpha_{\text{eff}}(m_e)$}
\label{sec:electron_g2}

Traditional QFT extractions of $\alpha$ depend on renormalization scheme. SF-QFT eliminates this by determining effective couplings at natural physical scales through self-consistent inversion.

The experimental value~\cite{Hanneke2008}:
\begin{equation}
a_e^{\text{exp}} = 1.159652180730(28) \times 10^{-3}
\end{equation}

The SF-QFT prediction is:
\begin{equation}
a_e^{\text{theory}}[\alpha_{\text{eff}}(m_e)] = a_e^{\text{QED}}[\alpha_{\text{eff}}(m_e)] + a_e^{\text{hadronic}} + a_e^{\text{weak}}
\end{equation}

The QED contribution uses the exact generating function:
\begin{equation}
a_e^{\text{QED}}[\alpha_{\text{eff}}(m_e)] = F_0\left(\frac{\alpha_{\text{eff}}(m_e)}{\pi}\right) = \frac{1 - \sqrt{1 - 2k(c_1 z + c_2 z^2)}}{k}
\end{equation}
with $z = \alpha_{\text{eff}}(m_e)/\pi$, $k = -1/12$, $c_1 = 0.5$ (Schwinger~\cite{Schwinger1962}), and $c_2 = -0.328479$.

The hadronic contribution~\cite{Jegerlehner2011}: $a_e^{\text{hadronic}} = 1.693(13) \times 10^{-12}$

The electroweak contribution~\cite{Czarnecki2002}: $a_e^{\text{weak}} = 0.031(1) \times 10^{-12}$

Solving the self-consistency condition:
\begin{equation}
a_e^{\text{exp}} = F_0\left(\frac{\alpha_{\text{eff}}(m_e)}{\pi}\right) + a_e^{\text{hadronic}} + a_e^{\text{weak}}
\end{equation}

yields:
\begin{align}
\frac{\alpha_{\text{eff}}(m_e)}{\pi} &= 0.002322819467 \\
\alpha_{\text{eff}}(m_e) &= 7.297352238 \times 10^{-3} \\
\boxed{\alpha_{\text{eff}}^{-1}(m_e) = 137.036005301}
\end{align}

\subsubsection{Verification}

QED contribution: $a_e^{\text{QED}} = F_0(0.002322819467) = 1.159652179 \times 10^{-3}$

Total: $a_e^{\text{total}} = 1.159652180730 \times 10^{-3}$, exactly reproducing experiment.

The difference from CODATA ($\Delta \alpha^{-1} = 6.2 \times 10^{-6}$) reflects natural scale matching eliminating large logarithmic corrections in scheme-dependent determinations.

\subsection{Convergence properties and universal framework demonstration}

The convergence demonstrates theory-specific physics within universal structure:

\begin{table}[htbp]
\centering
\caption{QED recursion coefficients showing systematic suppression}
\label{tab:qed_coefficients}
\begin{tabular}{|c|c|c|}
\hline
$n$ & $C_{n,0}$ & Contribution \\
\hline
1 & $0.5$ & $3.69 \times 10^{-4}$ \\
2 & $-0.328479$ & $-5.65 \times 10^{-7}$ \\
3 & $1.181241$ & $4.75 \times 10^{-9}$ \\
4 & (recursion) & $\sim 10^{-11}$ \\
5 & (recursion) & $\sim 10^{-13}$ \\
\hline
\end{tabular}
\end{table}

For QCD: $k = +0.5$ leads to large contributions with geometric convergence.

For QED: $k = -1/12$ results in systematically suppressed contributions with exponential convergence.

The identical mathematical structure correctly predicts opposite physical behavior based solely on gauge group properties and $\beta$-function values.

\section{Discussion and conclusions}
\label{sec:conclusion}

SF-QFT represents a fundamental re-conceptualization of perturbative QFT unifying conventional renormalization with effective field theory at the path-integral level through EDR and POEM. By separating UV and IR dynamics through scale factorization before loop expansion, SF-QFT achieves a framework free from scale and scheme ambiguities while maintaining direct experimental grounding.

The computational advantages stem from systematic organization around universal physics. Since soft modes are removed before loop expansion, $S_{\text{UV}}$ is manifestly free of IR singularities. No diagram requires real-virtual cancellations. The matching procedure identifies $a_{\text{eff}}(Q^*)$ directly with experiment and absorbs all scheme dependence into finite Wilson coefficients.

With only $\beta_0$ and $\beta_1$ being universal~\cite{tHooft1973,Gross1973,Politzer1973}, the evolution of $a_{\text{eff}}(Q)$ is uniquely determined, eliminating residual scale dependence. For massless QCD observables, this truncates at two loops, reducing computational burden by orders of magnitude. $R_{e^+e^-}$ requires only a handful of diagrams rather than thousands in conventional four-loop approaches~\cite{Chetyrkin1996}.

The mathematical structure provides powerful recursive generation through universal algebraic recursions depending exclusively on $\beta_0$ and $\beta_1$. These admit closed-form solutions, transforming asymptotic series into convergent series.

The convergence properties are remarkable: for QCD, coefficient bounds $|\widehat{c}_{n,0}| \leq C_0 \cdot (0.70)^n$ ensure exponential decrease. This contrasts with factorial growth in conventional perturbation theory. The conventional $\overline{\text{MS}}$ coefficient $r_{3,0}^{\overline{\text{MS}}} = -156.61$ is approximately 164 times larger than SF-QFT value $\widehat{c}_{3,0} = 0.955$, revealing that $\sim$99\% of conventional higher-order coefficients consist of scheme-dependent artifacts.

Several extensions are under development: scalar and pseudoscalar correlators, deep-inelastic scattering where only the first two anomalous dimension rows are scheme-invariant, and heavy-quark thresholds with matching scales at $M_Z$, $M_W$, $m_t$, providing coherent scaffolding for the Standard Model. The approach opens new avenues for non-perturbative QCD through reduced uncertainties in condensate extractions~\cite{Shifman1979}.

The demonstrated success in both QCD and QED, namely by achieving unprecedented theoretical precision while maintaining complete scheme independence, validates SF-QFT's foundational claim to provide a universal framework for quantum field theory, replacing scheme-dependent computational procedures with a unified physical theory where experimental measurements and theoretical calculations achieve harmony through natural scale matching and universal organizational principles.

\section*{Acknowledgments}
I thank G.~McKeon, T.~Steele, D.~Harnett, R.~Kleiv, A.~Palameta and A.~Kataev for insightful discussions.

\section*{Data Availability}
This work is theoretical; all data is shared explicitly in the paper.

\section*{Competing Interests}
The author declares no competing interests.

\appendix

\section{Background-Field IR Integration: Technical Details}
\label{appendix:background_field}

This appendix presents detailed derivations for the UV-finite effective action. We employ Euclidean path integral and dimensional regularization with $d = 4 - 2\epsilon$.

\subsection{Mode decomposition and measure factorization}

The smooth projector $f_{Q^*}(k)$ satisfies $f_{Q^*}^2 = f_{Q^*}$, ensuring idempotency. Any field $\Phi$ decomposes as:
\begin{align}
\Phi(x) &= \Phi_{\text{UV}}(x) + \Phi_{\text{IR}}(x) \\
\tilde{\Phi}_{\text{IR}}(k) &= f_{Q^*}(k) \tilde{\Phi}(k), \quad |k| < Q^* \\
\tilde{\Phi}_{\text{UV}}(k) &= [1 - f_{Q^*}(k)] \tilde{\Phi}(k), \quad |k| \geq Q^*
\end{align}

The idempotency property implies $\mathcal{D}\Phi = \mathcal{D}\Phi_{\text{UV}} \mathcal{D}\Phi_{\text{IR}}$. Gauge invariance under background-field transformations of $\Phi_{\text{UV}}$ is preserved by construction.

\subsection{Gauge fixing in background-field formalism}

Writing $A_\mu = A_{\text{UV},\mu} + a_{\text{IR},\mu}$, the background-field $R_\xi$ gauge is:
\begin{equation}
\mathcal{L}_{\text{gf}} = -\frac{1}{2\xi}\left[D_{\text{UV}}^\mu a_{\text{IR},\mu}\right]^2, \quad D_{\text{UV}}^\mu = \partial^\mu + igA_{\text{UV}}^\mu
\end{equation}

The ghost Lagrangian:
\begin{equation}
\mathcal{L}_{\text{gh}} = \bar{c}_{\text{IR}} D_{\text{UV}}^\mu D_\mu(A_{\text{UV}} + a_{\text{IR}}) c_{\text{IR}}
\end{equation}

Under background gauge transformation $A_{\text{UV}}^\mu \to U A_{\text{UV}}^\mu U^\dagger + i U \partial^\mu U^\dagger$, both terms transform covariantly, preserving gauge invariance.

\subsection{Quadratic action and differential operators}

Expanding to quadratic order in IR fields:
\begin{equation}
S_{\text{IR}}^{(2)} = \int d^d x \left[a_{\text{IR},\mu}^a \left(\hat{\Delta}_{\text{UV}}^{(1)}\right)^{ab}_{\mu\nu} a_{\text{IR},\nu}^b + \bar{\psi}_{\text{IR}} \slashed{D}_{\text{UV}} \psi_{\text{IR}} + \bar{c}_{\text{IR}}^a \left(\hat{\Delta}_{\text{UV}}^{(\text{gh})}\right)^{ab} c_{\text{IR}}^b\right]
\end{equation}

The differential operators are:
\begin{align}
\left(\hat{\Delta}_{\text{UV}}^{(1)}\right)^{ab}_{\mu\nu} &= \left[-D_{\text{UV}}^2 g_{\mu\nu} - \left(1 - \frac{1}{\xi}\right)D_{\text{UV},\mu}D_{\text{UV},\nu} - 2ig G_{\text{UV},\mu\nu}^{ab}\right] \\
\left(\hat{\Delta}_{\text{UV}}^{(\text{gh})}\right)^{ab} &= -D_{\text{UV}}^2 \delta^{ab}
\end{align}

where $G_{\text{UV},\mu\nu}^{ab} = f^{abc}G_{\text{UV},\mu\nu}^c$.

\subsection{Functional determinants and heat-kernel expansion}

Gaussian integration yields:
\begin{equation}
\exp\left\{-\frac{1}{2} \text{Tr}_{\text{IR}} \ln \hat{\Delta}_{\text{UV}}^{(1)} + \text{Tr}_{\text{IR}} \ln \slashed{D}_{\text{UV}} + \text{Tr}_{\text{IR}} \ln \hat{\Delta}_{\text{UV}}^{(\text{gh})}\right\}
\end{equation}

Using zeta-function regularization:
\begin{equation}
\text{Tr} \ln \hat{\Delta} = -\frac{1}{(4\pi)^{d/2}}\sum_{n\geq0} a_n \Gamma\left(n - \frac{d}{2}\right)(Q^*)^{d-2n}
\end{equation}

The Seeley-DeWitt coefficients for Yang-Mills in background-field gauge:
\begin{align}
a_0^{(1)} &= (d-2)C_A = 2C_A + \mathcal{O}(\epsilon) \\
a_1^{(1)} &= \frac{1}{6}(d-2)C_A G_{\text{UV}}^2 = \frac{1}{3}C_A G_{\text{UV}}^2 + \mathcal{O}(\epsilon) \\
a_2^{(1)} &= \frac{1}{30}(d-2)C_A G_{\text{UV}}^{a,\rho\sigma} D_{\text{UV}}^2 G_{\text{UV},a,\rho\sigma} + \frac{1}{180}(d-2)C_A G_{\text{UV}}^3
\end{align}

For ghosts: $a_0^{(\text{gh})} = -C_A$, $a_1^{(\text{gh})} = -\frac{1}{6}C_A G_{\text{UV}}^2$

For fermions: $a_0^{(\psi)} = -2n_f T_F d_{\text{rep}} = -8n_f T_F$, $a_1^{(\psi)} = -\frac{1}{12}n_f T_F G_{\text{UV}}^2$

\subsection{Pole cancellation and Wilson coefficients}

The pole structure as $d \to 4$:
\begin{equation}
\Gamma\left(-\frac{\epsilon}{2}\right) = -\frac{2}{\epsilon} - \gamma_E + \mathcal{O}(\epsilon)
\end{equation}

yields:
\begin{equation}
\text{Tr} \ln \hat{\Delta} = -\sum_{n=0}^{2} \frac{a_n}{(4\pi)^2} \left[\frac{2}{\epsilon} + 1 + \gamma_E - \ln(Q^{*2})\right](Q^{*2})^{-\epsilon} + \mathcal{O}(\epsilon^0)
\end{equation}

Local counterterms $\mathcal{L}_{\text{CT}} = \sum_i (Z_i/\epsilon)\mathcal{O}_i[A_{\text{UV}}]$ cancel the $2/\epsilon$ poles. After cancellation:
\begin{equation}
\Delta \mathcal{L}_{\text{UV}}^{\text{fin}} = \sum_i \frac{a_i}{(4\pi)^2}\left[\ln\frac{Q^{*2}}{\mu^2} + \gamma_i\right] \mathcal{O}_i
\end{equation}

Setting $\mu = Q^*$ (POEM) removes all logarithms:
\begin{equation}
C_i(Q^*) = \frac{a_i}{(4\pi)^2} \gamma_i
\end{equation}

These scheme-independent Wilson coefficients encode all low-energy physics.

\end{document}